\documentclass[12pt]{article}
\usepackage[margin=21mm]{geometry}
\usepackage[authoryear,longnamesfirst]{natbib}
\usepackage{graphicx} % Required for inserting images
\usepackage{amsmath}
\usepackage{amsfonts,amsthm}
\usepackage{bm}

\begin{document}
\let\WriteBookmarks\relax
\def\floatpagepagefraction{1}
\def\textpagefraction{.001}
\newtheorem{theorem}{Theorem}
\newtheorem*{remark}{Remark}
\newtheorem{lemma}[theorem]{Lemma}
\newtheorem{proposition}[theorem]{Proposition}
\newtheorem{corollary}[theorem]{Corollary}
\title{Painlevé quadrics and their confluence for the dispersionless Kadomtsev-Petviashvili equation}
\newcommand{\pp}[1]{{p}_{\scriptscriptstyle \mathrm{#1}}}

\author{Benoit Huard$^{*}$ \and Lewis Patterson \and Anh-Asma Koheeallee\\[5mm]
${}^{*}$ School of Engineering, Physics and Mathematics, \\ Northumbria University, Newcastle upon Tyne, UK}
% \author{Benoit Huard}[orcid=0000-0003-2368-7493]
% \ead{benoit.huard@northumbria.ac.uk}
% \cortext[cor1]{Corresponding author}
% \cormark[1]
% \fnmark[1]
% \author{Lewis Patterson}
% \author{Anh-Asma Koheealle}

% \shortauthors{B. Huard et al.}
% \shorttitle{Painlev\'e quadrics of dispersionless KP}

%\affiliation[1]{organisation={Northumbria University},
%addressline={School of Engineering, Physics and Mathematics},
%country={UK}
%}

\def\tsc#1{\csdef{#1}{\textsc{\lowercase{#1}}\xspace}}
%\tsc{WGM}
%\tsc{QE}
\newcommand{\PI}{\mathrm{P}_{\mathrm I}}
\newcommand{\PII}{\mathrm{P}_{\mathrm{II}}}
\newcommand{\PXXXIV}{\mathrm{P}_{\mathrm{XXXIV}}}
\newcommand{\PItwo}{\mathrm{P}_{\mathrm I}^{2}}
\newcommand{\R}{\mathbb{R}}
\newcommand{\C}{\mathbb{C}}

\newcommand{\us}{u_{\scriptscriptstyle *}}

\newcommand{\dd}{\mathrm{d}}
\newcommand{\XX}{\bm{X}}
\newcommand{\rr}{\bm{r}}
\newcommand{\hh}{\bm{h}}

\date{October 2026}

\maketitle

\begin{abstract}
We construct and study a distinguished class of solutions of the dispersionless Kadomtsev--Pet\-viash\-vi\-li (dKP) equation whose geometry is governed by the first and second Painlev\'e equations.
The level surfaces of these solutions correspond to three geometric families of central quadrics, and we determine when their level surfaces degenerate into planes.
In the parabolic family parametrised by $\PI$, degenerate levels occur at the stationary points of the transcendent.
For the hyperbolic family parametrised by rational solutions of $\PII$, we show that degenerate levels occur at points where consecutive solutions intersect and we give a precise count of such degeneracies.
In the homogeneous elliptic family, the Hastings--McLeod solution of $\PII$ is characterised, up to sign, as the unique non-trivial real pole-free solution with Airy asymptotics whose quadric has no finite degenerate level.  
%Finally, we show that under the critical scaling associated with the rational solutions of Painlev\'e II, the rational $\PII$ quadrics converge to the parabolic tritronqu\'ee quadrics, with error $O(n^{-1/5})$.
Finally, under critical scaling, the quadrics generated by the $n$-th rational Painlevé-II solution converge locally, and away from poles, to the parabolic quadrics associated with the Painlevé-I tritronquée solution, with error $O(n^{-1/5})$.
\end{abstract}

 % \begin{quote}
 %     {\it You do seem to care about exact solutions.\\
 %     B. Dubrovin}
 % \end{quote}

\section{Introduction}
\noindent The dispersionless Kadomtsev--Petviashvili  equation
\begin{equation}
    (u_t-u u_x)_x-u_{yy}=0
    \label{eq:dKP}
\end{equation}
is one of the fundamental examples of an integrable partial differential equation of hydrodynamic type in \(2+1\) dimensions, and a prototype equation for multidimensional wavebreaking
\citep{manakov2011dispersionless, grava2016shock}.
Introducing a potential $v$ by $u_y=v_x$,
one may equivalently write (\ref{eq:dKP}) as the first-order system
\begin{equation}
    u_t-u u_x-v_y=0,\qquad u_y=v_x.
    \label{eq:dKP-system}
\end{equation}
This system arises as the dispersionless limit of the KP equation and plays a central role in the theory of multidimensional integrable systems. Its integrability may be characterized by the method of hydrodynamic reductions, according to which solutions are sought in terms of an arbitrary number of Riemann invariants satisfying compatible systems of hydrodynamic type \citep{ferapontov2004characterization,ferapontov2004integrability}.
A particularly interesting geometric class of solutions characterised by the requirement that the level surfaces $u=C \in \mathbb{R}$ are central quadrics in the $(x,y,t)$ space was obtained by
\citet{dunajski2002einstein} for (\ref{eq:dKP}) in the context of the twistor approach.
Such solutions are parametrised by the Painlev\'e I and II transcendents, while Painlev\'e III quadrics had been constructed by \citet{tod1995scalar} for the $\mathrm{SU}(\infty)$-Toda equation. 
%These solutions are characterised by the requirement that the level surfaces $u=C \in \mathbb{R}$ are central quadrics in the $(x,y,t)$ space. 
These constructions were placed in a broader setting by \citet{Ferapontov2012centralquad}. Using the method of hydrodynamic reductions, they classified the integrable models possessing the central quadric ansatz, obtaining five canonical forms that include the Boyer--Finley and dKP equations. 
Applying the ansatz to each canonical form produces all six Painlev\'e equations $\mathrm{P_{I}}$--$\mathrm{P_{VI}}$, with $\mathrm{P_{VI}}$ corresponding to the generic case.%, and, at least for dKP, the resulting solutions form a subclass of the two-phase solutions provided by hydrodynamic reductions. 

In this paper, we provide a geometric construction of central quadric solutions of (\ref{eq:dKP-system}) using the generalized hodograph method \citep{tsarev1991geometry}. 
In the two-component formulation \citep{grundland2007conditional}, one can introduce a potential $\Phi = z(u,v) t + m(u,v)$ \citep{manas2002reductions} which generates the hodograph transform
\begin{equation}
    x=\Phi_u, \quad y = \Phi_v.
    \label{eq:hodograph}
\end{equation}
For fixed values of \(u\) and \(v\), this parametrizes a straight line in \((x,y,t)\)-space with direction vector    $\rr=(z_u,z_v,1)$ and base point $ \hh=(m_u,m_v,0).$
Thus the generalized hodograph transform naturally associates solutions of (\ref{eq:dKP-system}) with two-parameter congruences of lines, and the central quadric ansatz \citep{dunajski2002einstein} imposes the additional condition that the level surfaces of \(u\) are central quadrics
\begin{equation}
    \XX^T M(u)\XX=\rho(u),\qquad \XX=(x,y,t)^T,
    \label{eq:central-quadric}
\end{equation}
where \(M(u)\) is a symmetric \(3\times 3\) matrix. 
For fixed $u$, it is then seen that the direction vector of the ruling $\XX=\hh+t \rr$ of the quadric lies on a projective conic. %into needs to satisfy the following conditions
% \begin{equation}
%     r^TMr=0,\qquad h^TMr=0,\qquad h^TMh=1.
%     \label{eq:quadric-line-conditions}
% \end{equation}
% The first of these conditions says that the direction vector \(r\) lies on a conic in projective direction space. 
%This director conic is the basic projective object organizing the reduction. Its type determines the Painleve equation which appears in the reduced dynamics.
%The aim of this paper is to revisit the central-quadric reductions of dKP from this projective line-congruence viewpoint. The compatibility of \ref{eq:hodograph} with the central-quadric condition leads to the constraint
%\begin{equation}
%    z_{vvv}=A z_v+B,
%    \label{eq:director-conic}
%\end{equation}
It is shown that the first Painlev\'e equation \(P_{\mathrm I}\) parametrises the parabolic case, while Painlev\'e II and XXXIV govern the non-parabolic cases. 
We show that degenerate quadric levels of (\ref{eq:central-quadric}) occur at the stationary points of the $\PI$ transcendent in the parabolic case, and at those of the momentum function of the  Painlev\'e II system \citep{okamoto1986studies} for the hyperbolic case.
%the $\PI$ transcendent in the parabolic case, and of the momentum function $\Theta$ of the Painlev\'e II system in the hyperbolic case \citep{okamoto1986studies}. 
These degenerate levels correspond to double planes.

The paper is organized as follows. 
In Section~\ref{sec:hodograph}, we recall the generalised hodograph construction for dKP, formulate it in terms of line congruences and derive the director-conic constraint.
In Section \ref{sec:quadrics}, we show how the parabolic, hyperbolic, and elliptic normal forms lead to the Painlev\'e reductions, and study particular congruences arising from distinguished solutions of the represented Painlev\'e equations, namely the tritronqu\'ee solution of $\PI$ and the Hastings--McLeod and rational solutions of $\PII$. 
% Section \ref{sec:Examples} is devoted to particular congruences arising from distinguished solutions of the represented Painlev\'e equations, namely the Hastings--McLeod solution of $P_{II}$ and the tritronquee solution of $P_I$.
Section~\ref{sec:largen} is devoted to the confluence of the rational quadrics at the real Vorob'ev--Yablonskii corner.%, and %Section~\ref{sec:conclusion} concludes.

\section{Generalised hodograph and line congruences}
\label{sec:hodograph}
\noindent 

The central-quadric ansatz has a simple interpretation in the hodograph variables.  
A solution of dKP determines a two-parameter family of straight
lines in $(x,y,t)$-space, and requiring these lines to lie on the level quadrics gives a set of incidence relations between the congruence and the quadratic form.  
These relations determine the ruled geometry of the level sets and the projective conic formed by their ruling directions, which will underlie the classification in the following sections.
Let
\begin{equation}
\Phi(u,v;t)=z(u,v)t+m(u,v).
\label{Hodograph-map}
\end{equation}
Under the hodograph transform (\ref{eq:hodograph}), system (\ref{eq:dKP-system}) becomes
\begin{eqnarray}
    &&z_{uu} - z_v z_{uv} + (z_u + u)z_{vv} = 0, \label{eq-z}
    \\
    &&m_{uu} - z_v m_{uv} + (z_u + u)m_{vv} = 0, \label{eq-m}
\end{eqnarray}
for every $t$ such that the map is invertible. 
Equation (\ref{eq-z}) first appeared in \citep{gibbons1996reductions} in the context of reductions of the Benney chain.
%The characteristic velocities of (\ref{eq-z}) 
The invertibility condition of the map (\ref{eq:hodograph}) is then given as
\begin{eqnarray}
    J&=&\frac{\partial (x,y)}{\partial (u,v)} = \left|\begin{array}{cc}\Phi_{uu}&\Phi_{uv} \\ \Phi_{uv} & \Phi_{vv}\end{array}\right|=\left|\begin{array}{cc}
    z_{uu} t+ m_{uu} & z_{u v} t + m_{u v} \\
    z_{u v} t + m_{u v} & z_{v v} t + m_{v v}
    \end{array}\right|\nonumber\\
    &&\quad = (z_{uu} z_{vv} - z_{uv}^2) t^2 +
    (z_{uu} m_{vv} + m_{uu} z_{vv} -2 z_{uv} m_{uv} ) t +  (m_{uu} m_{vv} - m_{uv}^2) 
    \neq 0.
    \label{Jacobian}
\end{eqnarray}
The gradient-catastrophe condition is precisely the focal-surface condition for the hodograph line congruence
\[
\det(\XX_u,\XX_v,\XX_t)=0.
\]
This is also equivalent to the $v$-independent envelope condition
\begin{equation}
    F(\XX,u) = \XX^T M(u) \XX-\rho(u) = 0, \quad F_u(\XX,u)= \XX^T \dot{M}(u) \XX -\dot{\rho}(u)= 0,
\end{equation}
where the dot denotes differentiation with respect to $u$, whenever $\nabla_{\mathbf X}F\ne0$ and $(\Phi_{uv},\Phi_{vv}) \neq (0,0)$. 
Where these conditions fail, the focal condition must be checked directly from \(J=0\). 
%In particular, a double-plane level need not consist entirely of focal points.
%Indeed, differentiating $F(X(u,v,t),u)=0$ with respect to $v$, $t$ and $u$ gives $F_X\cdot X_v=F_X\cdot X_t=0$ and $F_X\cdot X_u+F_u=0$. At regular points of the level quadric, where $F_X\neq0$ and $X_v$, $X_t$ span its tangent plane, $\det(X_u,X_v,X_t)$ therefore vanishes if and only if $F_X\cdot X_u=0$, that is, if and only if $F_u=0$.
%The central-quadric ansatz has a simple interpretation in the hodograph variables.  
For each fixed value of \(u\), the variable \(v\) parametrises a one-parameter family of straight lines in \((x,y,t)\)-space,
\[
\XX = (x,y,t)=\hh(u,v)+t\,\rr(u,v).
\]
Requiring that the level sets of $u$ be the central quadrics (\ref{eq:central-quadric}) and that the hodograph lines be contained in them yields the three incidence conditions
\begin{equation}\label{eq:incidence}
  \rr^{T}M\rr=0,\qquad \rr^{T}M\hh=0,\qquad \hh^{T}M\hh=\rho(u).
\end{equation}
Indeed, $(\hh+t\rr)^{T}M(\hh+t\rr)=\rho(u)$ for all $t$ if and only if
(\ref{eq:incidence}) holds. 
It is easily seen that the resulting non-degenerate level quadrics necessarily correspond to one-sheeted hyperboloids.
\begin{lemma}
\label{lem:ruled}
%Let $M$ be a real symmetric $3\times3$ matrix for which there exist $\rr,\hh\in\R^3$ satisfying (\ref{eq:incidence}). 
%Then $\det M\le 0$, with equality if and only if $M\rr=0$.
%Consequently every non-degenerate level set of a central-quadric dKP congruence is a one-sheeted hyperboloid, and the congruence
%is one of its two rulings.
Let \(\XX^TM(u)\XX=\rho(u)\) be a real central-quadric level containing a hodograph line \(\XX=\hh+t\rr\) with $\hh\neq 0$, and $\hh$ and $\rr$ are linearly independent. If \(M\) is non-degenerate, then \(\rho\neq0\) and \(\mathrm{sgn}(\rho)M\) has signature \((2,1)\). Hence the level is a one-sheeted hyperboloid. Moreover, \(\rho=0\) necessarily implies $\det{M}=0$.
\end{lemma}

\begin{proof}
Complete $\{\rr,\hh\}$ to a basis $\{\rr,\hh,\bm{k}\}$ of $\R^3$ and let $B$ be
the matrix with these columns. By (\ref{eq:incidence}),
\[
  B^{T}MB=
  \left(\begin{array}{ccc}
  0 			& 0 		&\alpha\\ 
  0 			&\rho(u)	&\beta\\ 
  \alpha 	&\beta	&\gamma
  \end{array}
  \right),
  \qquad \alpha=\rr^{T}M\bm{k},\ 
  \beta=\hh^{T}M\bm{k},\ 
  \gamma=\bm{k}^{T}M\bm{k},
\]
% whence $\det M=- \rho(u) \alpha^2/(\det B)^2$. Equality holds iff $\alpha=0$, i.e.
% iff $M\rr$ annihilates the whole basis. A real symmetric $M$ with $\det M<0$ has
% signature $(+,+,-)$ or $(-,-,-)$; the latter is excluded since
% $\hh^{T}M\hh=1>0$. A quadric $\XX^{T}M\XX=1$ with $M$ of signature $(+,+,-)$ is a
% one-sheeted hyperboloid, and it contains the line $\XX=\hh+t\rr$.
whence
\begin{equation}
(\det B)^2\det M
=
\det(B^TMB)
=
-\rho\alpha^2,
\end{equation}
and if $M$ is non-degenerate, then $\alpha\neq0$ and $\rho\neq0$.
%Suppose first that $\rho>0$. Then $\det M<0$, while
%$\hh^TM\hh=\rho>0$. Thus $M$ has signature $(2,1)$.
%If $\rho<0$, then $\det M>0$ and $\hh^TM\hh<0$, so $M$ has
%signature $(1,2)$; equivalently, $-M$ has signature $(2,1)$.
Therefore, 
\(
\mathrm{sgn}(\rho)\,M
\)
has signature $(2,1)$, since, for example $\rr$ is isotropic.
% \[
% \XX^T\bigl(\operatorname{sgn}(\rho)M\bigr)\XX=|\rho|.
% \]
The central quadric is thus a one-sheeted hyperboloid. 
Since it contains the line $\XX=\hh+t\rr$, the hodograph congruence is one of its rulings.
It is immediate that $\rho=0$ implies the level quadric is degenerate.
\end{proof}
\noindent 
As we shall see below, the quadrics described by Painlev\'e transcendents satisfy $\det M= C \rho(u)^2$ with a constant $C \neq 0$, so that $\rho=0$ is the necessary and sufficient condition for degeneracy.
For each fixed $u$, the hodograph lines form one ruling of the level quadric. 
According to the first equation in (\ref{eq:incidence}), their directions lie on a projective conic, 
\begin{equation}\label{eq:director}
  \mathcal{C}(u)\ :\quad
  \left(\begin{array}{ccc}  z_u & z_v & 1\end{array}\right) M(u)
    \left(\begin{array}{c}  z_u \\ z_v \\ 1\end{array}\right)=0,
\end{equation}
which we will refer to as the \emph{director conic}. 
In turn, the differential constraint
\begin{equation}\label{eq:constraint}
  z_{vvv}= a_1 z_v+ a_0,\qquad a_0,a_1\in\R,
\end{equation}
shown to be compatible with (\ref{eq-z}) by \citet{kaptsov2005linear}, thus amounts to selecting the type of $\mathcal{C}(u)$. 

Note that differentiating the first condition in (\ref{eq:incidence}) with respect to $v$ leads to $\rr^T M \rr_v = 0$. 
This implies that $\hh \in \mathrm{span}\{\rr,\rr_v\}$, and $\hh=\lambda(u,v) \rr_v$ as $h$ has vanishing third component. 
In turn, it can be easily seen from (\ref{eq-z}), (\ref{eq-m}) and (\ref{eq:constraint}) that $\lambda$ must be a constant which can be absorbed into the coordinate frame $(x,y,t)$.
Thus, we choose $\hh=\rr_v$ (i.e. $m=z_v$) directly into the reductions below without loss of generality.

\section{Painlev\'e quadrics}
\label{sec:quadrics}
\begin{figure}[h!]
    \centering
\includegraphics[width=0.95\linewidth]{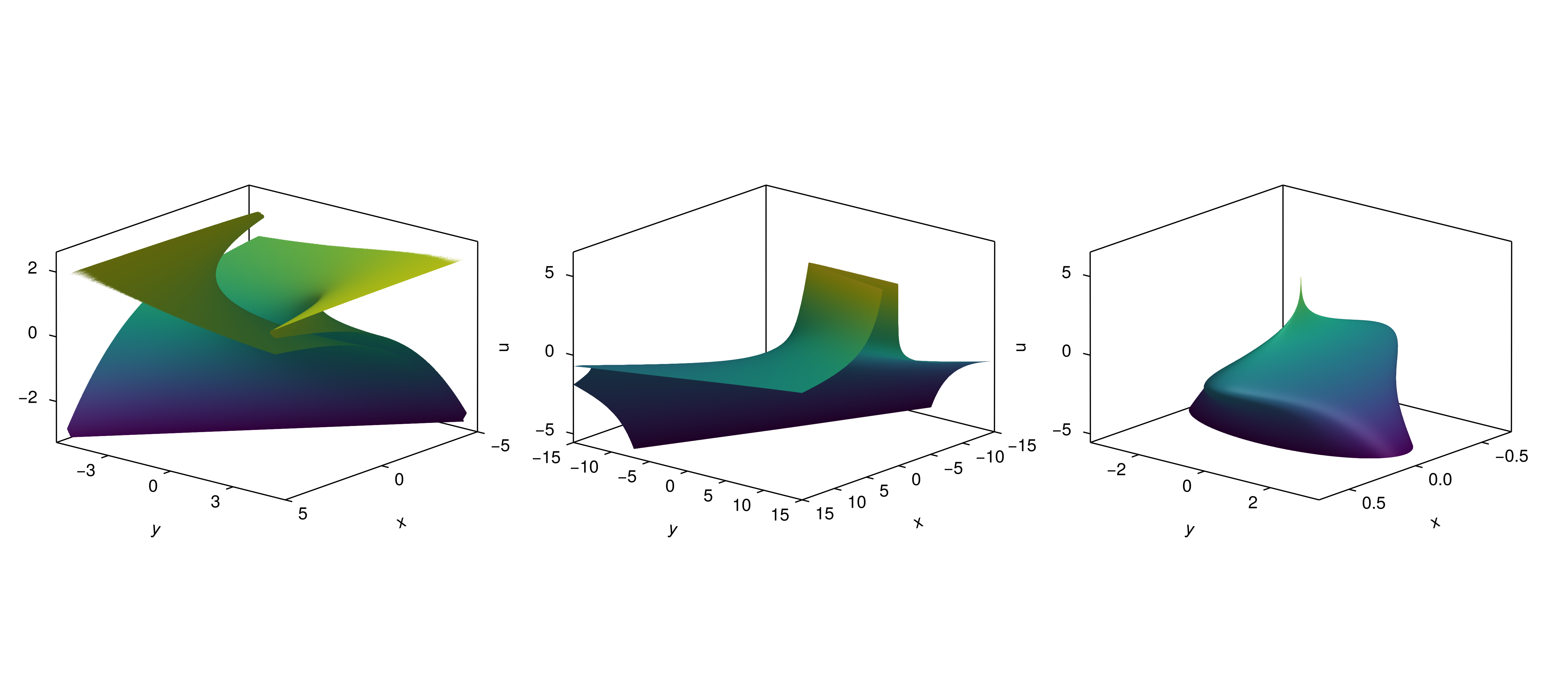}
    \caption{Central quadric solutions of dispersionless KP parametrised by the tritronqu\'ee solution of Painlev\'e I (left), and the Hastings-McLeod solution $q_{HM}$ of Painlev\'e II in the hyperbolic (centre) and elliptic (right) cases, in the $(x,y,u)$ space at $t=0$.}
    \label{fig:Surfaces}
\end{figure}

We now impose the differential constraint (\ref{eq:constraint}) and construct the reductions according to the affine type of the director conic $\mathcal{C}(u)$ in (\ref{eq:director}). 
For $a_1=0, a_0\neq0$ the conic is parabolic and the quadrics are governed by $P_{\rm I}$ (Section~\ref{sec:parabolic}); for $a_1>0$ it is hyperbolic and for $a_1<0$ elliptic, and both are governed by $P_{\rm II}$ or its equivalent form $P_{\rm XXXIV}$
(Sections \ref{sec:hyperbolic} and \ref{sec:elliptic}). 
In each case, we identify the degenerate levels where the quadric collapses to a double plane. These correspond to the stationary points of the relevant transcendent, and we study them for distinguished solutions: the tritronqu\'ee solution of $P_{\rm I}$, the Hastings--McLeod solution of $P_{\rm II}$, and the rational solutions of $P_{\rm II}$. 
Figure~\ref{fig:Surfaces} shows the resulting solution surfaces $u(x,y)$ at $t=0$.
% for the tritronqu\'ee (left) and Hastings--McLeod (hyperbolic, centre; elliptic, right) cases.

\subsection{Parabolic case}
\label{sec:parabolic}
\noindent Setting $a_1=0, a_0\neq0$, in (\ref{eq:constraint}) leads to the form
\begin{equation}
    z(u,v) = c_0(u) + c_1(u) v + c_2(u) v^2 + \kappa v^3, \quad \kappa \in \mathbb{R}\setminus \{0\},
    \label{z-P1}
\end{equation}
for which (\ref{eq-z}) is equivalent to the system
\begin{eqnarray}
&&c_0'' + 2 c_2 (c_0'+u)-c_1 c_1' =0, \quad  c_1'' + 6 \kappa  c_0'-2 c_1 c_2'+6 \kappa  u = 0, \quad c_2'' + 3\kappa  c_1'-2 c_2 c_2' = 0.
\label{sys-parabolic}
\end{eqnarray}
This system can be solved to give
\begin{eqnarray*}
    c_0'(u) &=& 2  \left[ \pp{I}(u)^2 +\pp{I}(u) \left(\int \pp{I}(u)du\right)^2-\pp{I}'(u) \int \pp{I}(u)du \right],\nonumber\\
    c_1(u) &=& \sqrt{2} \left(\left(\int \pp{I}(u) du\right)^2 - \pp{I}(u)\right), \quad   c_2(u) = \int \pp{I}(u) du,
\end{eqnarray*}
where we used the normalisation $\kappa = \left(3\sqrt{2}\right)^{-1}$, and $\pp{I}(u)$ satisfies the Painlev\'e I equation
\begin{equation}
\pp{I}''(u) = 6 \pp{I}(u)^2 + u.
\label{Painleve1}
\end{equation}
% Note that these expressions can also be written integral-free thanks to the first integral
% $$\int p(u) du = 2 p(u)^3 + u p(u) - \frac{1}{2} p'(u)^2.$$
The corresponding ruling in hodograph form reads as
\begin{eqnarray*}
    x &=& \left(v^2 \pp{I}(u) + \sqrt{2} v \left(2 \pp{I}(u) \int \pp{I}(u)du-\pp{I}'(u)\right) +2 \pp{I}(u) \left(\int \pp{I}(u)du\right)^2 -2 \pp{I}'(u) \int \pp{I}(u)du+2 \pp{I}(u)^2\right)t \nonumber \\
   &&\quad +2 v \pp{I}(u)+ \sqrt{2}\left( 2  \pp{I}(u) \int \pp{I}(u)du- \pp{I}'(u)\right), \\
   y &=& \left(\frac{v^2}{\sqrt{2}} + 2 v \int \pp{I}(u)du+ \sqrt{2} \left(\int \pp{I}(u)du\right)^2- \sqrt{2} \pp{I}(u)\right) t + 2 \int \pp{I}(u)du+\sqrt{2} v
\end{eqnarray*}
Eliminating $v$ leads to the central quadric
\begin{equation}
   X^T M(u) X = 2 \pp{I}'(u)^2, 
\end{equation}
with 
\begin{equation}
    M(u) = \left(
\begin{array}{ccc}
 1 & -\sqrt{2} \pp{I}(u) & -4 \pp{I}(u)^2 \\
 -\sqrt{2} \pp{I}(u) & 2 \pp{I}(u)^2 & \sqrt{2} \left(4 \pp{I}(u)^3-\pp{I}'(u)^2\right) \\
 -4 \pp{I}(u)^2 & \sqrt{2} \left(4 \pp{I}(u)^3-\pp{I}'(u)^2\right) & 4 \pp{I}(u) \left(4 \pp{I}(u)^3-\pp{I}'(u)^2\right) \\
\end{array}
\right),
\end{equation}
%For $t=0$, this quadric reduces to the pair of lines
%$$x - \sqrt{2} p(u) y = \pm \sqrt{2} p'(u).$$
which can be written in the factorised form
\begin{equation}
    \left(x-\sqrt{2} \pp{I}(u) y - 4 \pp{I}(u)^2 t\right)^2 = 2 \pp{I}'(u)^2 \left( 1+ \sqrt{2} y t + 2 \pp{I}(u) t^2\right).
    \label{P1-quadric}
\end{equation}
One checks that $\det M=-2\pp{I}'(u)^4=-\frac12\rho^2$, with $\rho=2\pp{I}'(u)^2$.
Hence, degenerate levels of the parabolic family are characterised by the stationary points of the corresponding Painlev\'e-I transcendent.
For the real tritronqu\'ee \citep{Boutroux1913}, however, no such degeneracy occurs on the real pole-free branch.
% \begin{remark}
%     The case $\kappa=0$ in (\ref{z-P1}) directly leads to
%     \begin{equation}
%         c_0''+2 c_2 c_0'-c_1 c_1'+2 u c_2=0, \quad c_1'' - 2 c_1 c_2' = 0,  \quad c_2' - c_2^2 = \alpha \in \mathbb{R}.
%     \end{equation}
% \end{remark}

\begin{proposition}
Let $p_{\rm tri}$ denote the real tritronqu\'ee solution of
\[
p''=6p^2+u,
\qquad
p(u) \sim -\sqrt{-u/6}
%-\frac{1}{48u^2}+O\bigl(|u|^{-9/2}\bigr)
\qquad u\to-\infty,
\]
and let $u_*\approx 2.384 > 0$ be its first real pole. 
Then
\[
p_{\rm tri}'(u)>0,\qquad u \in (-\infty, u_*),
\]
so that the associated parabolic quadrics are non-degenerate on this
interval. Moreover, there exists a unique stationary point between any two consecutive real poles. 
At such levels, the corresponding quadric (\ref{P1-quadric}) degenerates to a double plane.
\end{proposition}
\begin{proof}
The real tritronqu\'ee is strictly monotonic up to its first pole, as proved by \citet[Theorem 4]{JoshiKitaev2001} (after reversing the real axis).
Hence $\pp{tri}'(u)>0$ for $u \in (-\infty,u_{*})$. 
Furthermore, for $u>0$ and away from poles,
\[
p_{\rm tri}''(u)=6p_{\rm tri}(u)^2+u>0.
\]
%so $p_{\rm tri}'$ is strictly increasing on every pole-free interval.
%In particular, $p_{\rm tri}'$ remains positive on $(0,u_*)$, proving the first statement.
Every pole, say $u_j>0$, of a Painlev\'e-I solution is a double pole with local
behaviour \citep{ince1956ordinary,JoshiKitaev2001}
\[
p(u)=\frac{1}{(u-u_j)^2}-\frac{u_j}{10}(u-u_j)^2+O\bigl((u-u_j)^3\bigr),
\qquad
p'(u)=-\frac{2}{(u-u_j)^3}+O(u-u_j).
\]
Since all real poles of the tritronqu\'ee solution are positive \citep{CostinHuangTanveer2014}, $\pp{tri}'$ is continuous and strictly increasing on each interval $(u_j,u_{j+1})$ between consecutive positive poles, from $-\infty$ to $+\infty$. 
It thus possesses exactly one zero there, and the parabolic quadric (\ref{P1-quadric}) reduces to a double plane at each stationary point of $\pp{tri}$.
\end{proof}

% \noindent The invertibility of the hodograph map leads to a caustic that is a cubic polynomial in $v$, with discriminant
% \begin{eqnarray}
%    \tilde{J}(u,t) &=&-8 t^2 p'(u)^2 \Big[27 t^6 p'(u)^4-2 \left(6 t^2 p(u)-1\right) \left(6 t^2 p(u) \left(3 t^2 p(u)+2\right)+9 t^4 u-1\right) p'(u)^2  \nonumber\\
%    &&\hspace{4cm} +t^2 \left(6 p(u)^2+u\right)^2
%    \left(12 t^2 p(u) \left(t^2 p(u)+1\right)+8 t^4 u-1\right)\Big]
% \end{eqnarray}
\noindent For fixed $(u,t)$, the focal condition $J=0$ in (\ref{Jacobian}) is generically a cubic equation in $v$ with discriminant
\begin{equation}\label{J-caustic}
  \tilde J(u,t)=-8\,t^2\,p'(u)^2\,\Delta(u,t),
\end{equation}
where
\begin{align*}
  \Delta(u,t)={}&27t^6p'^4-2\bigl(6t^2p-1\bigr)\bigl(6t^2p\,(3t^2p+2)+9t^4u-1\bigr)p'^2 \\
  &\qquad +t^2\bigl(6p^2+u\bigr)^2\bigl(12t^2p\,(t^2p+1)+8t^4u-1\bigr).
\end{align*}

\begin{figure}
    \centering
    \includegraphics[width=0.95\linewidth]{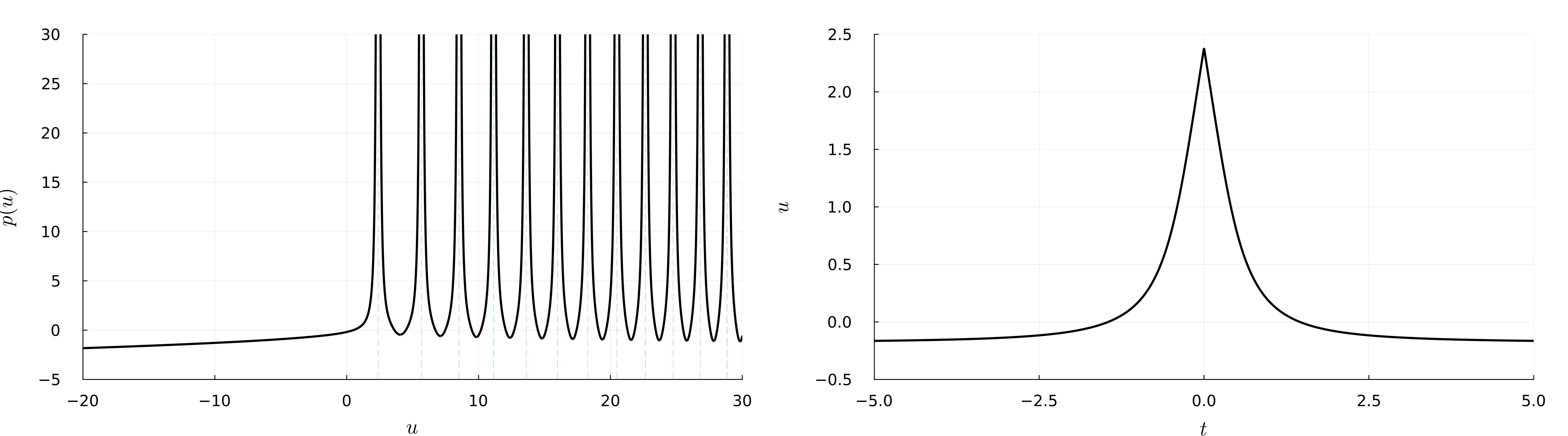}
    \caption{Tritronqu\'ee solution of Painlev\'e I, with the first pole located at $u=u_*\approx 2.384$ (left), and the curve $\tilde{J}(u,t)=0$ on which the discriminant (\ref{J-caustic}) vanishes for $t\neq0$, in the $(t,u)$ domain up to the first pole (right).}
    \label{fig:tritronquee-caustic}
\end{figure}
The prefactor vanishes only at $t=0$ and at the stationary points of $p$, where the level quadric collapses to a double plane. %since the tritronqu\'ee has none for $u<u_*$, the region $\tilde J>0$ of three focal lines is bounded there by the curve $\Delta=0$ alone (Figure~\ref{fig:tritronquee-caustic}).
%At each real pole $u_0$, the curve pinches to a corner with slopes $du/dt=\pm\sqrt{27/2}$.
\begin{proposition}\label{prop:corner}
Let $u_0$ be a real pole of a solution of \eqref{Painleve1}. 
The two nontrivial branches of the discriminant locus approaching \((t,u)=(0,u_0)\) are asymptotic to
\begin{equation}\label{eq:corner}
 t=\pm\sqrt{\frac{2}{27}}\,(u-u_0),\qquad
 \left|\frac{\dd u}{\dd t}\right|\longrightarrow\sqrt{\frac{27}{2}}.
\end{equation}
%For the tritronqu\'ee, $u_0=u_*$ is therefore the endpoint of the regular real caustic.
\end{proposition}
\begin{proof}
Put $\delta=u-u_0$ and $t=\tau\delta$. The Laurent expansion at a $\PI$
pole gives
\[
 p=\delta^{-2}+O(\delta^2),\qquad
 p'=-2\delta^{-3}+O(\delta),\qquad
 6p^2+u=6\delta^{-4}+O(1).
\]
Substitution into \eqref{J-caustic} yields
\[
 \widetilde J=\frac{-128\tau^2(27\tau^2-2)}{\delta^{10}}+O(\delta^{-6}).
\]
Thus $\widetilde J=0$ gives $\tau^2=2/27$, proving \eqref{eq:corner}.
\end{proof}

\subsection{Hyperbolic case}
\label{sec:hyperbolic}
If $a_1=1$ (and $a_0=0$), setting
\begin{equation}
    z(u,v) = s_1(u) + s_2(u) e^v + s_3(u) e^{-v}
    \label{eq:zhyp}
\end{equation}
leads to the system
\begin{equation}
    s_1'' + 2 (s_2 s_3)' = 0,\quad 
    s_2'' + (s_1' + u)s_2 = 0,\quad
    s_3'' + (s_1' + u)s_3 = 0. \label{hyp-sys}
\end{equation}
The general solution can be written as
\begin{equation}\label{eq:ssol}
  s_1'=-\Theta,\qquad
  s_2=\frac{\Theta}{2}e^{-\int q\,\dd u},\qquad
  s_3=e^{\int q\,\dd u},
\end{equation}
where 
\begin{equation}\label{eq:T}
  \Theta(u)=u+q(u)^2+q'(u),
\end{equation}
is Okamoto's momentum function for Painlev\'e II \citep{okamoto1986studies}.
Here, the Wronskian of $s_2, s_3$ is constant, 
\(
  s_2 s_3'-s_3 s_2'=-c,
\)
and \(q\) satisfies
\begin{equation}\label{eq:qeq}
  q''=2q^3+2uq+2c-1.
\end{equation}
Using (\ref{eq:qeq}), one obtains the identities 
\begin{eqnarray}\label{eq:Thetaid}
  &&\Theta'=2(q\Theta+c),\\
  %\Theta''=\frac{\Theta'^2}{2\Theta}
  %+2\Theta^2-2u\Theta-\frac{2c^2}{\Theta}.
  \label{eq:Theta-PXXXIV}
  &&2 \Theta \Theta''=\Theta'^2
  +4\Theta^3-4u\Theta^2-4c^2,
\end{eqnarray}
showing that \(\Theta\) satisfies the Painlev\'e XXXIV equation
\citep{ince1956ordinary}.
Setting $q(u)=2^{1/3}\pp{II}(s)$ with $s=2^{1/3}u$ transforms (\ref{eq:qeq}) into the second Painlev\'e equation
\begin{equation}\label{eq:PII}
  \pp{II}''(s)=2 \pp{II}(s)^3+s \pp{II}(s)+\alpha,\qquad \alpha=c-\frac12 .
\end{equation}
% The conserved quantity $s_2 s_3' - s_3 s_2' = c$ allows one to obtain the solution
% \begin{equation*}
%    \hspace{-5mm} s_1(u) = -\frac{u^2}{2} - q(u) - \int q^2(u)\,du, \, 
%     s_2(u) 
% =
% \frac{q'(u)+q(u)^2+u}{2}
% \exp\left(-\int q(u)\,du\right),\,
%     s_3(u) = \mathrm{exp}\left(\int q(u)\right) du,
% \end{equation*}
% where $q(u)$ satisfies
% \[
% q''=2q^3+2 u q + 2c-1.
% \]
% In turn, introducing $q(u) = 2^{1/3} p(s), s=u^{1/3}$, leads to the Painlev\'e II equation
% \[
% p''(s)= 2p(s)^3 + s p(s) + c - \frac{1}{2}.
% \]
The resulting hodograph equations
\begin{eqnarray*}
x&=&t \left(\frac{1}{2} \left(2 c+q(u) \left(q'(u)+u\right)+q(u)^3\right) e^{v-\int q(u) \, du}-q'(u)+q(u) e^{\int q(u)
   \, du-v}-q(u)^2-u\right)\nonumber \\
   &&+\frac{1}{2} \left(2 c+q(u) \left(q'(u)+u\right)+q(u)^3\right) e^{v-\int q(u) \, du}-e^{\int q(u) \, du-v} q(u),\\
y&=&t \left(\frac{1}{2} \left(q'(u)+q(u)^2+u\right) e^{v-\int q(u) \, du}-e^{\int q(u) \, du-v}\right)+\frac{1}{2}
   \left(q'(u)+q(u)^2+u\right) e^{v-\int q(u) \, du}+e^{\int q(u) \, du-v},
\end{eqnarray*}
provide, upon elimination of $v$, the central quadric solution
\begin{equation}
    X^T M X = -2(q \Theta+c)^2 = -\frac12 \Theta'^2, 
    \label{hyperbolic-quadric}
\end{equation}
with
\[
M(u) = \left(
\begin{array}{ccc}
 \Theta & -c & \Theta^2 \\
 -c & -q (2 c+q \Theta) & -c \Theta \\
 \Theta^2 & -c \Theta & \Theta^3-2 q^2 \Theta^2 -4 c q \Theta -2 c^2 \\
\end{array}
\right).
 \]
One checks that $\det M=2(q\Theta+c)^4=\frac12\rho^2$.
Using identity (\ref{eq:Thetaid}), the central quadric can also be written in the equivalent form
\begin{equation}
    \left(x+\Theta t - \frac{c}{\Theta} y\right)^2 = \frac{\Theta'^2}{4 \Theta^2} (y^2 + 2 \Theta (t^2 - 1)),
    \label{hyperbolic-square-1}
\end{equation}
for any $u$ such that $\Theta(u)\neq 0$.\\
The case \(c=0\) is exceptional, whereby (\ref{eq:Thetaid}) reduces to \(\Theta'=2q\Theta\). Hence, on any interval where
\(\Theta\not\equiv0\), the function \(\Theta\) cannot vanish, and the
hyperbolic quadric simplifies to
\[
(x+\Theta t)^2-q^2y^2 - 2q^2\Theta t^2 = -2q^2\Theta.
\]
Its degenerate levels therefore occur precisely at zeros of \(q\), or
equivalently of \(\Theta'\), and at such a point \(u=u_d\) the quadric
collapses to the double plane
\[
(x+\Theta(u_d)t)^2=0.
\]
The exceptional branch \(\Theta = q'+q^2+u=0,  \forall u \in \mathbb{R}\) corresponds to the Riccati reduction of Painlev\'e II at \(\alpha=-\tfrac12\), and is equivalent to the Airy equation \citep{flaschka1980monodromy}. 
This case leads to the vanishing of the constraint (\ref{hyperbolic-quadric}), and therefore does not provide a central quadric solution. 

% \begin{equation}\label{eq:hypcornerfact}
%    \left(x+\Theta t+q y\right)\left(\Theta (x+\Theta t)-(2c+q\Theta)y \right)=\frac12 \Theta'^2 (t^2-1),
% \end{equation} 
% or, after using identity (\ref{eq:Thetaid}).
% The normal form
% \begin{equation}
%      t^2+\frac{y^2}{2\Theta}
%  -\frac{2\Theta}{\Theta'^2}
%  \left(x+\Theta t-\frac{c}{\Theta}y\right)^2
%  =1
%  \label{hyp-normal}
% \end{equation}
% is valid for $u \in \mathbb{R}$ such that $\Theta(u) \Theta'(u) \neq 0$. 
For $c\neq 0$, degenerate levels can be characterised through the zeros of $\Theta'$, for which the quadric (\ref{hyperbolic-square-1}) degenerates to a double plane. 
This can be established more precisely for the Hastings--McLeod solution of Painlev\'e II as well as for its rational solutions.
\begin{proposition}
\label{prop:HMhyperbolic}
Let $q_{\scriptscriptstyle \rm HM}$ be the positive Hastings--McLeod solution of
\begin{equation}
q''=2q^3+2uq,
\label{HM-theo}
\end{equation}
obtained from the standard homogeneous $P_{\scriptscriptstyle \mathrm{II}}$ solution by rescaling.
Let us denote
\[
q_\pm=\pm q_{\scriptscriptstyle \rm HM},\qquad
\Theta_\pm=u+q_\pm^2+q_\pm'.
\]
For $c=\frac{1}{2}$ the associated hyperbolic quadrics (\ref{hyperbolic-quadric}) satisfy \(\rho=-\frac12\Theta_\pm'^2.\) 
The family associated with $q_-$ is non-degenerate for every finite $u$,
whereas the family associated with $q_+$ has exactly one finite degenerate
level. At this level, the hyperbolic quadric collapses to a double plane.
\end{proposition}

\begin{proof}
For $c=\frac12$, the identity (\ref{eq:Thetaid}) becomes
\begin{equation}
\Theta'=2q\Theta+1,
\label{idenTheta-1}
\end{equation}
for any solution $q(u)$ of the homogeneous Painlev\'e $\mathrm{II}$ equation.
For the positive Hastings--McLeod solution, we obtain the asymptotics as $u\to-\infty$
\[
q_+(u)\sim\sqrt{-u},\quad 
\Theta_+(u)\sim-\frac{1}{2\sqrt{-u}},
\quad
\Theta_+'(u)\sim-\frac{1}{4(-u)^{3/2}}<0.
\]
As $u\to+\infty$, $q_+$ has Airy decay, so $\Theta_+'(u)\to 1$ and thus $\Theta_+'$ has at least one zero. 
At any zero of $\Theta'$, identity (\ref{idenTheta-1}) and the facts that $q_+ > 0$ and $q_+'<0$ imply
\begin{equation}
    \Theta_+ = -\frac{1}{2 q_+}<0, \quad \Theta'' = 2 q_+' \Theta_+ > 0.
\end{equation}
Hence every zero of $\Theta_+'$ is a crossing from negative to positive values, so that there exists a unique $u_+$ such that $\Theta_+'(u_+) = 0$.
% is a crossing from negative to positive, so there can be only one. 
By the hyperbolic normal form (\ref{hyperbolic-square-1}), the degenerate level is a double plane
\[
\left(x+ q_+(u_+) y - \frac{c}{q_+(u_+)} t\right)^2 = 0.
\]
For $q_-=-q_{\scriptscriptstyle \rm HM}$, the behaviour at $u\to-\infty$ is
\[
q_-(u)\sim -\sqrt{-u}, \quad 
\Theta_-(u)\sim\frac{1}{2\sqrt{-u}},
\qquad
\Theta_-'(u)\sim\frac{1}{4(-u)^{3/2}}>0,
\]
while again $\Theta_-'(u)\to1$ as $u\to+\infty$.
At any hypothetical zero of $\Theta'_{-}$, $q_-<0$, $q_-'>0$ and
\[\Theta_-=-\frac{1}{2q_-} >0, \quad \Theta_-''=2q_-'\Theta_->0.\]
Since $\Theta_-'$ is already positive for large negative $u$, such a zero cannot occur. 
Thus $\Theta_-'(u)>0$ for all finite $u$ and the associated central quadric has no degenerate level.
%%%Figure
%\begin{figure}
%    \centering
%    \includegraphics[width=0.75\linewidth]{hastings_mcleod_hyperbolic-zero.png}
%    \caption{The positive Hastings--McLeod solution \(q_+\) of (\ref{HM-theo}) for \(c=\frac12\),
%and the associated functions
%\(\Theta_+=u+q_+^2+q_+'\) and
%\(\Theta_+'=2q_+\Theta_++1\).
%The unique zero \(u_d\) of \(\Theta_+'\) determines the unique finite
%degenerate level.}
%    \label{fig:placeholder}
%\end{figure}
\end{proof}
\subsubsection{Rational Painlev\'e II solutions}
The rational solutions of the Painlev\'e II equation in the form
\begin{equation}
    q_n'' = 2 q_n^3 + 2 u q_n + 2 n, \quad n \in \mathbb{N},
    \label{q-PII-n}
\end{equation}
are given as
\begin{equation}
    q_n(u) = 2^{1/3} \frac{d}{ds} \ln \left(\frac{Q_{n-1}(s)}{Q_n(s)}\right)\Bigg|_{s=2^{1/3} u},
    \label{qn-voro}
\end{equation}
where the Vorob'ev-Yablonskii polynomials $Q_n$ are defined by the following recursion formula
\begin{equation}
Q_0 = 1, \quad Q_1(s) = s,\quad Q_{n+1}(s)Q_{n-1}(s) = sQ_n^2(s) - 4 \left(Q_n(s)Q_n''(s) - Q_n'(s)^2\right), \quad n \in \mathbb{N}.
\end{equation}
The scheme is easily extended to negative integers $n$ using the symmetry $(q,n)\to(-q, -n)$.
Numerous studies have focused on characterising the location and properties of zeros of these polynomials. %see for example \citep{clarkson2003second,clarkson2006special, roffelsen2012number,buckingham2015large}. 
\citet{fukutani2000special} proved that the zeros of every $Q_n$ are simple and that adjacent Vorob'ev--Yablonskii polynomials have no common zeros.  
In particular, it follows that the poles of $q_n$ are precisely the (rescaled) zeros of $Q_{n-1} Q_n$: a zero of $Q_{n-1}$ gives a simple pole of residue $+1$, while a zero of $Q_n$ gives a simple pole of residue $-1$.
The variable
\[
 \Theta_n=u+q_n^2+q_n'
\]
has particularly useful representations \citep{clarkson2006special} in terms of Vorob'ev--Yablonskii polynomials,
\begin{equation}\label{eq:ThetaVY}
 \Theta_n(u)
 =
 2^{-1/3}
 \frac{Q_{n+1}(s) Q_{n-1}(s)}
      {Q_n(s)^2} \Bigg|_{s=2^{1/3}u} = u-2\frac{d^2}{du^2}\log Q_n(2^{1/3}u).
\end{equation}
Formula (\ref{eq:ThetaVY}) immediately separates the different roles played by the real zeros of the three neighboring Vorob'ev--Yablonskii polynomials.
It implies that $\Theta_n$ vanishes at the zeros of $Q_{n-1}$ and $Q_{n+1}$, which are all simple and distinct from those of $Q_n$ \citep{clarkson2006special,roffelsen2012number}. 
The second equation in (\ref{eq:ThetaVY}) implies the asymptotics
\begin{equation}
\Theta_n(u) = u  + O(u^{-2}), \quad u\to \pm \infty.
\label{Theta-n-asymp}
\end{equation}
In turn, taking the logarithmic derivative away from the zeros of $\Theta_n$ yields
\begin{equation}
\frac{\Theta_n'(u)}{\Theta_n(u)}= q_n(u) - q_{n+1}(u).
\label{Theta-q}
\end{equation}
Thus, the degenerate levels of the hyperbolic quadric associated with the solution (\ref{qn-voro}) occur at the points of intersection between consecutive rational solutions
\begin{equation}
q_n(u) = q_{n+1}(u).
\label{q-intersections}
\end{equation}
The B\"acklund transform for Painlev\'e II
\begin{equation}
    q_{n+1}(u) = - q_n(u) - \frac{2n+1}{\Theta_n},
    \label{Backlund-PII}
\end{equation}
leads to the equivalent formulation of (\ref{q-intersections})
\[q_{n+1}(u) = q_{n}(u) \quad \leftrightarrow \quad 2 q_n(u) \Theta_n(u) = -(2n+1).\]

Let us show that there are exactly $\left\lfloor \frac{n+1}{2} \right\rfloor$ real roots for that condition. 
This corresponds to the number of real roots of the Vorob'ev-Yablonski polynomial $Q_n$, as demonstrated by \citet{roffelsen2012number}.
We first demonstrate that $\Theta_n$ does not have a critical point before its first pole.
\begin{lemma}
    Let $a_n$ be the left-most real pole of $q_n$ with residue $-1$. Then $\Theta_n'(u) > 1$ for $u \in (-\infty, a_n)$.
    \label{Lemma:Theta-n-zeros}
\end{lemma}
\begin{proof}
This can be shown by induction.
Let us assume that for \(n\geq1\), \(q_n\) has a first real pole \(a_n\) with residue \(-1\), and that
\[
q_n^{(k)}(u)>0,
\qquad
k=0,1,2,\ldots,
\]
for all $u \in (-\infty,a_n)$.
This pole is then also the first real pole of \(\Theta_n\).
The assertion clearly holds for $n=1$, in which case $q_1=-1/u$, $a_1=0$ and $\Theta_1 = u+2u^{-2}$ has a critical point at $u=4^{1/3}>0$.\\
Suppose now that it holds for \(q_n\). 
This implies that, on the one hand,
\[
\Theta_n'=1+2q_nq_n'+q_n''>1,
\]
and that every derivative of \(\Theta_n\) of positive order is strictly positive on \((-\infty,a_n)\). 
In turn, the asymptotics of $\Theta_n$ at $u\to-\infty$ given in (\ref{Theta-n-asymp}) imply that there exists a unique $b \in (-\infty, a_n)$ such that $\Theta_n(b) = 0$, and $\Theta_n < 0$ on $(-\infty,b)$. \\
% \[
% \Theta_n(u)\longrightarrow-\infty
% \quad (u\to-\infty),
% \qquad
% \Theta_n(u)\longrightarrow+\infty
% \quad (u\uparrow a_n),
% \]
% there is exactly one zero \(b<a_n\), and
% \(\Theta_n<0\) on \((-\infty,b)\).
On the other hand, using the differential equation (\ref{q-PII-n}), one can rewrite the B\"acklund transformation (\ref{Backlund-PII}) as
\[
q_{n+1}
=
\left(n+1+q_nq_n'+\frac12q_n''\right)
\left(-\frac1{\Theta_n}\right).
\]
Since the derivatives of each of these factors are strictly positive on $(-\infty, b)$, we obtain that
% Every derivative of the first factor is positive.
% Every derivative of the second factor is positive on
% \((-\infty,b)\) as well. Indeed, differentiating
% \[
% (-\Theta_n)\left(-\frac1{\Theta_n}\right)=1
% \]
% gives, for \(k\geq1\),
% \[
% \left(-\frac1{\Theta_n}\right)^{(k)}
% =
% \frac1{-\Theta_n}
% \sum_{j=1}^{k}
% \binom{k}{j}\Theta_n^{(j)}
% \left(-\frac1{\Theta_n}\right)^{(k-j)}>0,
% \]
% by induction on \(k\). The product rule therefore yields
\[
q_{n+1}^{(k)}(u)>0, 
\qquad k\geq0,  \quad u\in (-\infty,b).
\]
Moreover, \(\Theta_n'(b)=2n+1\), so the Bäcklund transformation implies
\[
q_{n+1}(u)=-\frac1{u-b}+O(1) \quad 
\text{as $u\to b$}.
\]
Thus \(a_{n+1}=b\) is the first real pole of \(q_{n+1}\) with residue
\(-1\), and is also the first real pole of
\(\Theta_{n+1}\). 
This completes the induction.
\end{proof}

\noindent We can now prove the following theorem.
\begin{theorem}
\label{theo-rational}
For each integer \(n\geq0\), let \(q_n\) be the rational
solution of
\[
q_n''=2q_n^3+2u q_n+2n,
\]
and set
\[
\Theta_n=u+q_n^2+q_n'.
\]
Then \(q_n\) and \(q_{n+1}\) have exactly r=\(
\left\lfloor\frac{n+1}{2}\right\rfloor
\)
finite real intersections, all of which are transverse crossings. 
At such points, the central quadric (\ref{hyperbolic-square-1}) reduces to double planes.
\end{theorem}

\begin{proof}
%Due to identity (\ref{idenTheta-1}), $\Theta_n$ and $\Theta_n'$ cannot vanish together. 
For \(n=0\), the assertion follows from \(q_0=0\), which is the unique rational solution in the homogeneous case, and
\(q_1=-1/u\). 
Henceforth assume \(n\geq1\). 
The B\"acklund transform (\ref{Backlund-PII}) and the differential equation give
\[
q_{n+1}=-q_n-\frac{2n+1}{\Theta_n},
%\qquad
%\Theta_n'=2q_n\Theta_n+2n+1.
\]
Furthermore, $\Theta_n$ satisfies the $P_{\scriptscriptstyle\mathrm{XXXIV}}$ equation,
\begin{equation}\label{eq:theta-second}
2\Theta_n\Theta_n''
=(\Theta_n')^2+4\Theta_n^3
 -4u\Theta_n^2-(2n+1)^2,
\end{equation}
which, upon differentiation yields
\begin{equation}\label{eq:theta-third}
\Theta_n'''=(6\Theta_n-4u)\Theta_n'-2\Theta_n.
\end{equation}
Equation (\ref{eq:theta-second}) implies that at a zero of \(\Theta_n\), we have
\[
\Theta_n' =\pm (2n+1) \neq 0.
\]
Thus \(\Theta_n\) and \(\Theta_n'\) cannot vanish together.
Consequently,
\[
q_n-q_{n+1}=\frac{\Theta_n'}{\Theta_n},
\qquad
q_n=\frac{\Theta_n'-(2n+1)}{2\Theta_n},
\qquad
q_{n+1}=-\frac{\Theta_n'+(2n+1)}{2\Theta_n},
\]
implying the finite intersections of \(q_n\) and \(q_{n+1}\) correspond exactly to the real stationary points of \(\Theta_n\).

By equation (\ref{eq:ThetaVY}), $\Theta_n$ has a double pole at each rescaled real root of $Q_n$, hence $\Theta_n$ has exactly $r=\left\lfloor\frac{n+1}{2}\right\rfloor$ real poles. 
At any pole $a$, $\Theta_n(u)$ has an expansion
\[\Theta_n(u) = \frac{2}{(u-a)^2} + O(1), \quad u\to a,\]
while 
\begin{equation}
\Theta_n(u)=u+O(u^{-2}), \qquad u\to\pm\infty.
\label{Theta-asymptotics}
\end{equation}
By Lemma \ref{Lemma:Theta-n-zeros}, no critical point of $\Theta_n$ can occur before its first real pole. 
We now prove that \(\Theta_n\) has exactly one critical point between consecutive real poles and after the last real pole. Let \(I\) be any one of these intervals. At both endpoints of \(I\), including \(+\infty\) when appropriate, \(\Theta_n\) tends to \(+\infty\). 

\noindent Equation \eqref{eq:theta-third} implies
\begin{equation}
\frac{d}{du}
\left(\Theta_n''-3\Theta_n^2+4u\Theta_n\right)
=2\Theta_n.
\label{Theta_diffs}
\end{equation}
Hence, if $u_1,u_2 \in I$ are such that $u_1<u_2$ and
\(\Theta_n(u_1)=\Theta_n(u_2)=\theta_0\neq0\), integration of (\ref{Theta_diffs}) and equation \eqref{eq:theta-second} give
\begin{equation}\label{eq:theta-equal-height}
\Theta_n'(u_2)^2-\Theta_n'(u_1)^2
=
4\theta_0\int_{u_1}^{u_2}\bigl(\Theta_n(u)-\theta_0\bigr)\,du.
\end{equation}
Let us use (\ref{eq:theta-equal-height}) to show that any critical point in $I$ is necessarily a non-degenerate minimum, and thus necessarily the global minimum over the interval.

Denoting \(\us \in I\) a critical point over $I$ and writing \(\theta_0=\Theta_n(\us)\neq0\), we claim that \(\Theta_n''(\us)>0\). 
Suppose instead that \(\Theta_n''(\us)\leq 0\). 
At this point, \eqref{eq:theta-third} gives
\[
\Theta_n'''(\us)=-2 \theta_0,
\]
leading to the Taylor expansion
\begin{equation}
\Theta_n(u) = \theta_0 + \frac{\Theta_n''(\us)}{2}(u-\us)^2 - \frac{2 \theta_0}{3!} (u-\us)^3 + O\left((u-\us)^4\right).
\label{Taylor-Theta}
\end{equation}
Under the assumption \(\Theta_n''(\us)\leq 0\), we then obtain the two cases.
\begin{itemize}
    \item {\bf $\theta_0>0$:} The Taylor expansion (\ref{Taylor-Theta}) shows that \(\Theta_n(u)<\theta_0\) immediately to the right of \(\us\). 
    Since \(\Theta_n\) tends to \(+\infty\) at the right endpoint of \(I\), there is a first \(\bar{u}>\us\) such that \(\Theta_n(\bar{u})=\theta_0\). 
    Thus
\[
\Theta_n(u)<\theta_0, \quad u \in (\us,\bar{u}).
\]

Equation \eqref{eq:theta-equal-height} now gives
\[
0\leq\Theta_n'(\bar{u})^2
=
4\theta_0\int_{\us}^{\bar{u}} \bigl(\Theta_n(u)-\theta_0\bigr)\,du<0,
\]
which is a contradiction.
\item {\bf \(\theta_0<0\):} Similarly, (\ref{Taylor-Theta}) implies that
\(\Theta_n(u)<\theta_0\) immediately to the left of \(\us\).
The limit \(+\infty\) at the left endpoint of \(I\)
therefore gives a last \(\bar{u}<\us\) such that
\(\Theta_n(\bar{u})=\theta_0\), with
\[
\Theta_n(u)<\theta_0, \quad u \in (\bar{u},\us).
\]
Applying \eqref{eq:theta-equal-height} on \([\bar{u},\us]\)
yields
\[
0\geq-\Theta_n'(\bar{u})^2
=
4\theta_0\int_{\bar{u}}^{\us} \bigl(\Theta_n(u)-\theta_0\bigr)\,du>0,
\]
which again leads to a contradiction.
\end{itemize}
Hence, every critical point in \(I\) satisfies $\Theta_n''(\us)>0$, and is thus a nondegenerate local minimum which must therefore be unique over $I$ as no local maximum can exist. 
Therefore, each interval $I$ contains a unique critical point of $\Theta_n$, providing $\lfloor\frac{n+1}{2}\rfloor$ crossing points between $q_n$ and $q_{n+1}$. 
Transversality is easily established by noting that
$$(q_n - q_{n+1})' = \frac{\Theta_n''}{\Theta_n} \neq 0.$$
This completes the proof.

%. The endpoint limits ensure that a global minimum exists in \(I\). There cannot be two critical points, since two strict local minima would force an intervening local maximum. Thus \(I\) contains exactly one critical point, which is its unique global minimum.

% At the corresponding intersection \(u_*\),
% \[
% (q_n-q_{n+1})'(u_*)
% =
% \frac{\Theta_n''(u_*)}{\Theta_n(u_*)}\neq0.
% \]
% Therefore every intersection is a transverse crossing.

% There are \(r-1\) intervals between consecutive real poles
% and one interval after the last pole. Each contributes
% exactly one crossing, while the interval before the first
% pole contributes none. 
% The total number of crossings is therefore
% \[
% (r-1)+1=r
% =\left\lfloor\frac{n+1}{2}\right\rfloor.
% \]
\end{proof}

\subsection{Elliptic case}
\label{sec:elliptic}
If $a_1=-1, a_0=0$ in (\ref{eq:constraint}), we set
\begin{equation}
    z(u,v) = s_1(u) + s_2(u) \cos{v} + s_3(u) \sin{v}
\end{equation}
and obtain the system
\begin{eqnarray}
    &&s_1''(u) - \frac{1}{2}(s_2^2+s_3^2)' = 0, \,\, s_2''(u)-s_2(u) \left(s_1'(u)+u\right) = 0,\,\, s_3''(u)-s_3(u) \left(s_1'(u)+u\right) = 0.
\end{eqnarray}
Introducing polar coordinates in the form
\[s_2 = 2 r(u) \cos{\phi(u)}, \quad s_3 =2 r(u) \sin{\phi(u)}\]
implies that 
\begin{eqnarray}
   r {\phi''}+2 {r' \phi'}=0, \quad 
r'' =  \left(u + 2 r^2+\phi '^2\right) r.
\end{eqnarray}
Since, as before, the Wronskian is preserved, we can write
\begin{equation}
    s_2 s_3' - s_3 s_2' = 4 r^2 \phi' = L \in \mathbb{R},\label{wronskian-elliptic}
\end{equation}
so that $\phi' = \frac{L}{4 r^2}$. 
Hence, the radial function $r(u)$ satisfies
\begin{equation}
r''\ =  2 r^3 + u r + \frac{L^2}{16 r^3}.
\label{Painleve2-r}
\end{equation}
Equivalently, $R=r^{2}$
satisfies the Painlev\'e XXXIV equation \citep{ince1956ordinary},
$$R'' = \frac{R'^2}{2R} + 4 R^2 + 2 u R + \frac{L^2}{8 R}.$$
If $L=0$, then (\ref{Painleve2-r}) is the homogeneous Painlev\'e II equation.
In turn, (\ref{wronskian-elliptic}) implies $\phi$ corresponds to a constant phase shift which can be set to $0$.
The resulting hodograph solution is then written as
\begin{equation}
    x=2 t \left(\cos (v) r'(u)+r(u)^2\right)-2 \sin (v) r'(u),\quad y= -2
   t r(u) \sin (v) - 2 r(u) \cos (v),
   \label{rulings-elliptic}
\end{equation}
where $r(u)$ satisfies
\begin{equation}
r''\ =  2 r^3 + u r.
\label{Painleve2-homo}
\end{equation}
%Solving (\ref{rulings-elliptic}) for $\cos{v}$ and $\sin{v}$ and using the identity $\cos{v}^2+\sin{v}^2=1$ 
Eliminating $v$ in (\ref{rulings-elliptic}) leads to the central quadric solution $X^T M X = 4 r^2 r'^2$ with
\begin{equation} 
M=\left(
\begin{array}{ccc}
r^2 & 0 & -2 r^4 \\
0 & r'^2 & 0\\
-2 r^4 & 0 & 4r^2 (r^2-r')(r^2+r')
\end{array}
\right),
    \label{quad-dkp-elliptic}
\end{equation}
or equivalently for any value of $u$ where $r(u) r'(u) \neq 0$,
\begin{equation}\label{eq:ellnormal}
  \frac{\left(x-2r^2t\right)^2}{(2r')^2}+\frac{y^2}{(2r)^2}-t^2=1.
\end{equation}
The real solutions of the homogeneous Painlev\'e-II equation (\ref{Painleve2-homo}) which are regular for large positive $u$ form the Airy family
\(
r(u)\sim k\,\mathrm{Ai}(u), 
\)
as $u\to+\infty$.
For $0<|k|<1$ these are the Ablowitz--Segur solutions
\citep{AblowitzSegur1977} which are pole-free on $\R$ and oscillatory about zero as
$u\to-\infty$.  
The endpoint values $k=\pm 1$ give the Hastings--McLeod
solution and its negative \citep{hastings1980boundary}.
For $k=1$,
\[
r(u)\sim\mathrm{Ai}(u),\qquad u\to+\infty,
\qquad
r(u)\sim\sqrt{-u/2},\qquad u\to-\infty,
\]
with $r>0$ and $r'<0$ on $\R$.  For $|k|>1$ the corresponding real
solution develops poles on the real axis.  
%Thus decay at $+\infty$ does not by itself distinguish Hastings--McLeod; what is special is its global pole-freeness together with the absence of finite zeros of both $r$ and $r'$.
We thus obtain the following proposition.
\begin{proposition}
    Let $r(u)$ be a nontrivial real solution of (\ref{Painleve2-homo}), regular on $\mathbb{R}$, such that $\lim_{u\to\infty} r(u)=0.$
The family of central quadrics (\ref{eq:ellnormal}) is non-degenerate for every finite $u \in \mathbb{R}$ if and only if $|k|=1$, i.e. $r(u)$ is, up to sign, the Hastings-McLeod solution.  
%Thus, up to the symmetry $r\mapsto-r$, the Hastings--McLeod solution is the unique non-trivial real pole-free Airy solution whose quadrics (\ref{eq:ellnormal}) have no finite degenerate level.
\end{proposition}
\begin{proof}
For $|k|=1$, the Hastings--McLeod solution and its negative satisfy
$r(u)\neq0$ and $r'(u)\neq0$ for every finite $u$.  Hence
\[
\rho(u)=4r^2r'^2>0,
\]
so every finite level is non-degenerate. For $0<|k|<1$, the Ablowitz--Segur solution oscillates about zero as $u\to-\infty$ and therefore $\rho=4r^2r'^2$ vanishes infinitely often. The case $|k|>1$ is excluded by the occurrence of real poles \citep{deift1995asymptotics}.
\end{proof}

\section{Confluence at the real Vorob'ev--Yablonskii corner}
\label{sec:largen}

\begin{figure}[h!]
    \centering
    \includegraphics[height=50mm]{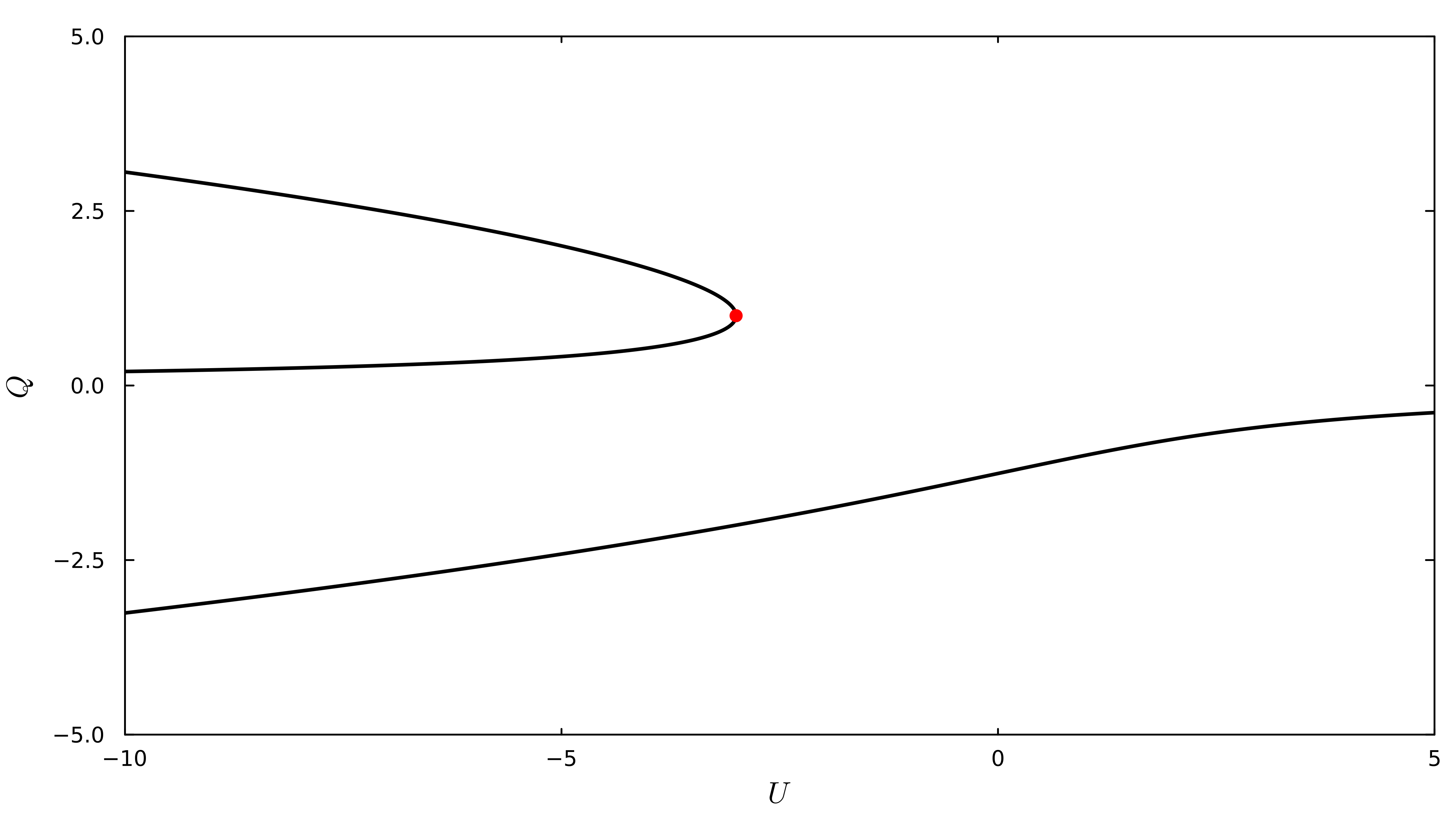}
    \includegraphics[height=50mm]{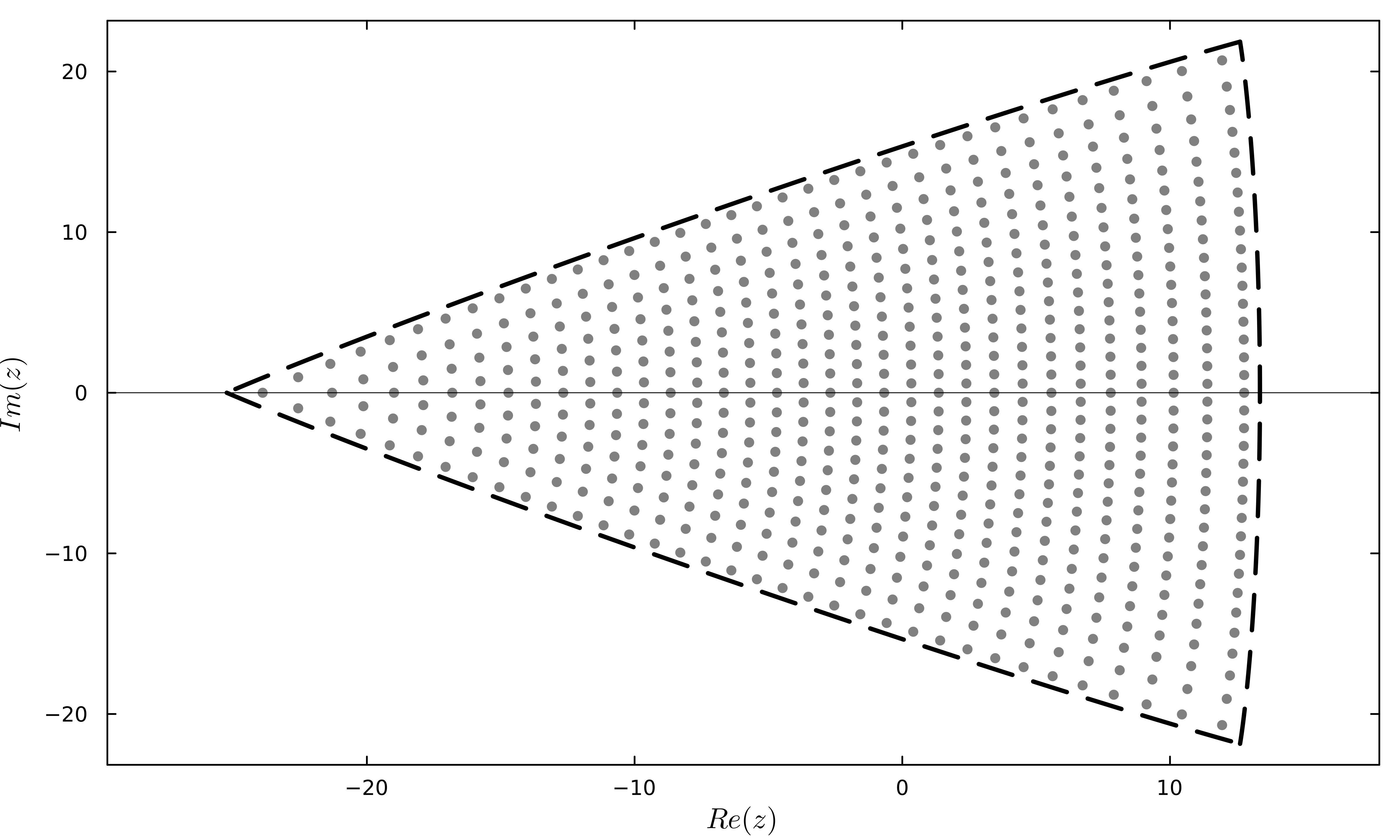}
    \caption{{Left}. The scaled outer cubic $Q^3+UQ+2=0$.  The branch followed by the rational $\PII$ solution as $U\to-\infty$ meets a second positive branch at the fold $(-3,1)$. {Right}. Scaled zeros of the Vorob'ev–Yablonskii polynomial \(Q_{35}\) in the complex \(z\)-plane (grey dots), together with the boundary of the associated curvilinear triangular region (dashed curve). As \(n\to\infty\), each vertex approaches a corner of opening angle \(2\pi/5\). Under the corresponding corner scaling, the rational Painlevé-II solution converges locally to a Painlevé-I tritronquée solution, uniformly on compact subsets away from its poles \citep{buckingham2015large}.} 
\label{fig:outercubic}
\end{figure}

For fixed \(n\), the Vorob'ev--Yablonskii polynomials determine the singular levels of the rational hyperbolic congruence. As \(n\to\infty\), their zeros fill a curvilinear triangular region in the scaled plane (Figure \ref{fig:outercubic}), whose three corners are critical points of the large-degree asymptotics
\citep{buckingham2014large,buckingham2015large, bertola2015zeros}.
We consider the real corner, where the critical scaling reduces $P_{\rm II}$ to $P_{\rm I}$, and the exponents are those of the $P_{\rm I}$ universality class observed at critical points of Hamiltonian PDEs \citep{dubrovin2009universality,DubrovinGravaKleinMoro2015}.
In this corner, the algebraic outer approximation to the rational \(P_{\mathrm{II}}\) solution develops a fold, and the corresponding critical rescaling converts \(P_{\mathrm{II}}\) into \(P_{\mathrm I}\).  
The same rescaling has a direct geometric effect on the dKP congruence: the hyperbolic director conic becomes parabolic and the hyperbolic level quadrics converge to the parabolic \(P_{\mathrm I}\) family.  
As shown by the Riemann--Hilbert analysis of 
\citet{buckingham2015large}, the rational Vorob'ev--Yablonskii family selects the tritronqu\'ee solution of Painlev\'e I at the real corner.

Indeed, consider again Painlev\'e II equation in the form \citep{buckingham2015large,halburd1996coalescence,kapaev1997scaling,kapaev1995passage}
\begin{equation}
q'' = 2 q^3 + 2 u q + 2 n, \quad n \in \mathbb{N},
\label{P2}
\end{equation}
where we have set $c=n+\frac{1}{2}$. Introduce the double scaling
\begin{equation}
u = u_c + \sigma \zeta, \quad q = A + \delta w(\zeta), \quad u_c, \sigma, \delta \in \mathbb{R}.
\label{corner-scaling}
\end{equation}
Equation (\ref{P2}) becomes
\begin{equation}
    \frac{\delta}{\sigma^2} \frac{d^2w}{d\zeta^2} = 2 (A^3 + A u_c + n) + 2 \delta (3A^2 + u_c) w + 6 A \delta^2 w^2 + 2 A \sigma \zeta + 2 \delta^3 w^3 + 2 \sigma \delta \zeta w.
    \label{P2-w}
\end{equation}
Requiring that (\ref{P2-w}) does not contain a linear nor a constant term in $w$ at $\zeta=0$ is equivalent to stating that 
\[u_c = -3A^2, \quad A = \left(\frac{n}{2}\right)^{1/3},\]
is a fold point of the cubic curve
$F(q,u) = q^3 +u q + n$.
Further introducing the variables $q = A Q, u = A^2 U$
leads to the algebraic cubic curve
\[Q^3 + U Q + 2 = 0,\]
for which the fold is located at $(-3,1)$, as illustrated in Figure \ref{fig:outercubic}. 
The further choice $\sigma = (2A^2)^{-1/5}$, $\delta = 2^{2/5} A^{-1/5}$ leads to
\begin{equation}
\frac{d^2w}{d\zeta^2} = 6 w^2 +  \zeta + \epsilon (2 w^3 + \zeta w),
\label{eq:coalres}
\end{equation}
where we have set $\epsilon = 2^{2/5} A^{-6/5}$.
\begin{remark}
Equivalently, introducing \(\xi=2/n=A^{-3}\), the critical scaling reads
\[
U+3=2^{-1/5}\xi^{4/5}\zeta,\qquad
Q-1=2^{2/5}\xi^{2/5}w,
\]
exhibiting the critical exponents characteristic of the classical Painlev\'e-I universality class  \citep{dubrovin2009universality,DubrovinGravaKleinMoro2015}.
\end{remark}
% The further choice $\sigma = (2A^2)^{-1/5}$, $\delta = 2^{2/5} A^{-1/5}$ leads to
% \begin{equation}
% \frac{d^2w}{d\zeta^2} = 6 w^2 +  \zeta + \epsilon (2 w^3 + \zeta w),
% \label{eq:coalres}
% \end{equation}
% where we have set $\epsilon = 2^{2/5} A^{-6/5}$. 
Let us now consider the effect of this scaling onto the hyperbolic quadric (\ref{hyperbolic-square-1}) written in the form
\begin{equation}
    \left(x+\Theta_n t - \frac{c}{\Theta_n} y\right)^2 = \frac{\Theta_n'^2}{4 \Theta_n^2} (y^2 + 2 \Theta_n (t^2 - 1))
    \label{hyperbolic-square}
\end{equation}
where $\Theta_n = u + q_n^2 + q_n'$, and $q_n$ is the rational solution as defined in (\ref{qn-voro}). 
In this case, the fold point $u_c$ corresponds to the real negative corner of the quasi-triangular region delimited by the zeros of the Vorob'ev-Yablonski polynomial $Q_n$ \citep[Proposition 1]{buckingham2015large}.  
Substituting the critical scaling (\ref{corner-scaling}) immediately gives, to leading order
\begin{equation}
\Theta_n = - 2 A^2 + O(A^{4/5}), \quad \Theta_n' = 2^{8/5} A^{6/5} \dot{w} + O(A^{3/5}).
\end{equation}
This implies that, to leading order,
\begin{equation}
y^2 + 2 \Theta_n t^2 = y^2 - 4 A^2 t^2 + O(A^{4/5} t^2). 
\end{equation}
% Hence,
% \begin{equation}
%     y^2 + 2 \Theta (t^2 - 1) = y^2 + 2 \Theta t^2 - 2 \Theta = (y-2 A t)(y+2 A t) + 2 (\Theta + 2 A^2 )t^2 - 2 \Theta
% \end{equation}
% Let us introduce a scaling 
% $$
% y - 2 A t = O(A^{m_1}), \quad y + 2 A t = O(A^{m_2})
% $$
% and balancing dominant terms implies $m_1+m_2=2$. 
% Furthermore, since $t=\frac{(y+2 A t)-(y - 2 A t)}{4A}$

% At the real corner, $\Theta=-2A^2+O(A^{4/5})$, so the quadratic factor
% $y^2+2\Theta t^2$ has leading null directions $y\pm2At=0$.
% Balancing the $O(A^{4/5})$ correction fixes
% \[
% y+2At=O(A^{2/5}),\qquad
% y-2At=O(A^{8/5}),\qquad
% t=O(A^{3/5}),
% \]
% which yields the stated $(Y,T)$ scaling.
% The $x$-scaling is then chosen to cancel the leading contribution
% $A(y-2At)$ in $x+\Theta t-(c/\Theta)y$, leaving a finite
% $O(A^{1/5})$ limit.

% ---
% The linear frame may be obtained directly from the null directions of
% the leading hyperbolic quadratic form.  Indeed, at the real corner
% \[
% \Theta=-2A^2+O(A^{4/5}),
% \]
% and hence
% \[
% y^2+2\Theta t^2
% =
% y^2-4A^2t^2+O(A^{4/5}t^2)
% =
% (y+2At)(y-2At)+O(A^{4/5}t^2).
% \]
\noindent The critical balance requires the constant term $-2\Theta_n=O(A^2)$,
the product of the two null coordinates, and the first correction
$(\Theta_n+2A^2)t^2$ to be of the same order.  This fixes
\[
y+2At=O(A^{2/5}),\qquad
y-2At=O(A^{8/5}),\qquad
t=O(A^{3/5}).
\]
We therefore introduce the normalized null coordinates
\[
y+2At=-2^{1/5}A^{2/5}Y,\qquad
y-2At=-2^{23/10}A^{8/5}T,
\]
or equivalently
\[
y=-2^{-4/5}A^{2/5}Y-2^{13/10}A^{8/5}T,\qquad
t=-2^{-9/5}A^{-3/5}Y+2^{3/10}A^{3/5}T.
\]
Furthermore,
\[
x+\Theta_n t-\frac{c}{\Theta_n}y
=
x+A(y-2At)+O(A^{1/5}),
\]
so the leading $O(A^{13/5})$ term can be cancelled by setting
\[
x=2^{1/10}A^{1/5}X+2^{23/10}A^{13/5}T.
\]
This yields the linear change of variables 
\begin{eqnarray}
\label{Smap}
 \hspace{-1cm}&&x=\mu X+\nu T, \quad y=a Y+b T, \quad t=f Y+g T,\\
 \hspace{-1cm}&&\mu =2^{1/10}A^{1/5},\, \nu=2^{23/10}A^{13/5},\, a=-2^{-4/5}A^{2/5}, \, b=-2^{13/10}A^{8/5},\, f=-2^{-9/5}A^{-3/5},\, g=2^{3/10}A^{3/5},\nonumber
 %x=2^{1/10}A^{1/5} X+2^{23/10}A^{13/5} T,\quad
 %&\qquad \mu&=2^{1/10}A^{1/5},& \nu&=2^{23/10}A^{13/5},\\
 %y=-2^{-4/5}A^{2/5}Y-2^{13/10}A^{8/5} T,\quad 
 %a&=-2^{-4/5}A^{2/5},& b&=-2^{13/10}A^{8/5},\\
 %t=-2^{-9/5}A^{-3/5} Y+2^{3/10}A^{3/5} T,
 %f&=-2^{-9/5}A^{-3/5},& g&=2^{3/10}A^{3/5}.
\end{eqnarray}
for which direct substitution leads to
\begin{eqnarray}
&&\frac{1}{\mu} \left(x+\Theta_n t - \frac{c}{\Theta_n} y\right) = X - \sqrt{2} w Y - 4 w^2 T + O(A^{-3/5})\\
&&\frac{1}{\mu^2} \left(\frac{\Theta_n'^2}{4 \Theta_n^2} \left(y^2 + 2 \Theta_n (t^2 - 1)\right)\right) = 2 \dot{w}^2 \left(1+\sqrt{2} T Y + 2 w T^2\right) + O(A^{-3/5})
\end{eqnarray}
Combined with the analysis of \citet{buckingham2015large}, which provides uniform asymptotic estimates showing that suitably scaled rational Painlevé-II solutions approach the Painlevé-I tritronquée solution on compact pole-free sets, we can thus establish convergence to the tritronqu\'ee parabolic quadric. 
The main results of \citep{buckingham2015large} can be summarised as follows, adapted to our notation.

\begin{lemma}[\cite{buckingham2015large}]
\label{lem:BM}
Let $p_{\rm tri}$ be the real tritronqu\'ee solution of
$p''=6p^2+\zeta$ fixed by $p_{\rm tri}(\zeta)=-\sqrt{-\zeta/6}+O(\zeta^{-2})$
as $\zeta\to\infty$, $|\arg(-\zeta)|\le 4\pi/5-\eta$, $0<\eta<\frac{4\pi}{5}$, and let $K\subset\mathbb{C}$ be compact and free of poles of $p_{\rm tri}$.
Set
\[
A=\Big(\tfrac{n}{2}\Big)^{1/3},\qquad
u_c=-3A^2,\qquad
\sigma=(2A^2)^{-1/5},\qquad
\delta=2^{2/5}A^{-1/5},
\]
and define
\[
w_n(\zeta)=\frac{q_n(u_c+\sigma\zeta)-A}{\delta},
\]
where $q_n$ is the rational solution of Painlev\'e II defined in (\ref{qn-voro}). 
Then, uniformly for $\zeta\in K$,
\[
w_n(\zeta)=p_{\rm tri}(\zeta)+O(n^{-1/5}), \quad \dot{w}_n=p_{\rm tri}'+O(n^{-1/5})\qquad n\to\infty .
\]
\end{lemma}
% \begin{proof}[Derivation from \cite{BuckinghamMiller2015}]
% In their notation $P_m$ solves
% $p_{yy}=2p^3+\tfrac23 yp-\tfrac23 m$, and $q_n(u)=-3^{1/3}P_n(3^{1/3}u)$
% solves \eqref{eq:PIIrational}. Their scaled variable is
% $x=(m-\tfrac12)^{-2/3}y$, with left-most corner
% $x_c=-(9/2)^{2/3}$, which corresponds to $u=u_c$ above.
% Their Theorem~3, eq.~(1.28), states that, uniformly for
% $t=2^{1/15}3^{-1/3}m^{4/5}(x-x_c)\in K$,
% \[
% m^{-1/3}P_m=-6^{-1/3}-m^{-2/5}\,2^{7/15}3^{-1/3}\,Y(t)+O(m^{-3/5}),
% \]
% where $Y=p_{\rm tri}$. Changing variables gives
% $q_n=A_n+\delta\,p_{\rm tri}(\zeta)+O(n^{1/3-3/5})$.
% Since $\delta\sim n^{-1/15}$, dividing by $\delta$ gives an error
% $O(n^{1/3-3/5+1/15})=O(n^{-1/5})$.
% \end{proof}
This finally leads to the theorem.
\begin{theorem}\label{thm:cornerconf}
Let $q_n$ be the rational solution of Painlev\'e $\mathrm{II}$ (\ref{P2}). 
Under the corner scaling (\ref{corner-scaling}) and the linear change of variables (\ref{Smap}), the hyperbolic director conic converges to the parabolic
director conic. 
On compact $\zeta$-sets avoiding poles of $\pp{tri}$, the corresponding level-quadric equation (\ref{hyperbolic-square}) converges to 
\begin{equation}
\label{eq:triquad}
 (X-\sqrt2\,\pp{tri} Y-4 \pp{tri}^2T)^2
 =2\pp{tri}'{}^2(1+\sqrt2\,Y T+2\pp{tri} T^2),
\end{equation}
with error $O(A^{-3/5})=O(n^{-1/5})$.
\end{theorem}

\begin{remark}
The map (\ref{Smap}) is also asymptotically compatible with dKP. Noting that \(
 \zeta=\frac{u-u_c}{\sigma}
\), a direct calculation gives
\begin{eqnarray*}
 (\zeta_T-\zeta\zeta_X)_X-\zeta_{YY}=
2^{-11/10}A^{-6/5}\zeta_{YT}
+
2^{-21/5}A^{-12/5}\zeta_{TT}=
O(A^{-6/5})=O(n^{-2/5}),
\end{eqnarray*}
matching the order of the perturbation in (\ref{eq:coalres}). 
The estimate thus holds uniformly on compact subsets of the \((X,Y,T)\)-domain where \(\zeta_{YT}\) and \(\zeta_{TT}\) remain uniformly bounded as \(A\to\infty\).
\end{remark}

\noindent {\bf Acknowledgments.} BH is grateful to E.V. Ferapontov, A. Moro, M. Sommacal, R. Dubertrand, R. Conte and M. Grundland for insightful discussions.\\ 

\noindent {\bf GenAI disclosure.} All calculations have been performed using Wolfram Mathematica 14.3.
Claude Opus 4.8 was used iteratively in the construction of the proof of Theorem \ref{theo-rational}. 
The authors have retained full responsibility for the writing of the document and the correctness of results.
\bibliographystyle{cas-model2-names}
\bibliography{Refs}

\end{document}